\documentclass[%
reprint,
superscriptaddress,
 amsmath,amssymb,
 aps,
]{revtex4-2}

\usepackage{float}

\usepackage{xcolor}

\usepackage{graphicx}
\usepackage{dcolumn}
\usepackage{bm}

\begin{document}

\preprint{APS/123-QED}

\title{ Tumbling elimination induced by permeability: an experimental approach}

\author{J. S\'anchez-Rodr\'iguez}
\affiliation{Dept. F\'isica Fundamental, Universidad Nacional de Educaci\'on a Distancia, Madrid, Spain}
\affiliation{Laboratory of Fluid Mechanics and Instabilities, \'Ecole Polytechnique F\'ed\'erale de Lausanne, CH-1015 Lausanne, Switzerland}

\author{F. Gallaire}%

\affiliation{Laboratory of Fluid Mechanics and Instabilities, \'Ecole Polytechnique F\'ed\'erale de Lausanne, CH-1015 Lausanne, Switzerland}

\date{\today}

\begin{abstract}
Archetypal falling behaviors of impervious objects are classified into four modes: fluttering, tumbling, steady descent and chaotic motion. The classical scenario predicts these behaviors to be affected by two dimensionless quantities: dimensionless inertia and Reynolds number. In this article we explore experimentally the effect of permeability and porosity on the falling regimes of porous plates. By drilling several hole distributions in rectangular plates, both permeability and porosity are varied systematically. We discover that the introduction of porosity affects the stability of the falling regimes eliminating tumbling. Using a phenomenological model we show that a decrease in circulation induced by the introduction of holes is the primary mechanism for stabilizing the plates' trajectories.
\end{abstract}

\maketitle

\onecolumngrid

\section{Introduction}

An everyday situation such as the fall of a leaf from a tree or a sheet of paper illustrates the intricate interaction between fluid and solid. Fluid-solid interaction of falling objects has received huge attention and has been studied from experimental, numerical and theoretical points of view \cite{willmarth1964steady, smith1971autorotating,  field1997chaotic,  pesavento2004falling, andersen2005unsteady}. Specifically, different regimes of falling objects in fluids have been identified and studied to determine  how they are influenced by the physical characteristics of both solid and fluid as well as the geometrical characteristics of the solid. The comprehensive understanding of this topic has diverse applications and encompasses fields as varied as ecology to aerodynamics \cite{ern2012wake}. These falling behaviors have been classically categorized into fluttering, tumbling, steady descent and chaotic motion. For thin objects, there are two dimensionless parameters involved in the selection of the falling regime: the dimensionless moment of inertia and the Reynolds number. Phase diagrams showing these behaviors have been elaborated for plates \cite{smith1971autorotating} or disks \cite{field1997chaotic}. The transition from one regime to the other therefore depends on these two dimensionless parameters, which in turn rely on physical and geometrical parameters.

However, in many circumstances, there are additional significant elements that may affect this interaction, like for instance the deformability of the structure. Porosity is another common trait that may be encountered in abundant situations. Some seeds, such as the common dandelion \cite{cummins2018separated}, are transported efficiently via a porous bundle of bristly filaments known as a pappus. Some insects have fringed or bristling wings instead of impermeable ones \cite{santhanakrishnan2014clap}. This configuration generates lift forces equivalent to smooth wings, but with a lower wing weight \cite{lee2020stabilized}. It is conceivable, thus, to include porosity and permeability and expand the study of falling object regimes to a broader framework, not only for a more realistic understanding of these natural phenomena, but also for their technical and industrial applications  \cite{kim2021three, chen2023light}. It is interesting to consider the potential stabilizing impact that porosity and permeability may provide to a falling object throughout its descent. In this context, trajectory stability occurs when the object's landing and release sites belong to the same vertical plane. In terms of trajectories, this implies an elimination of tumbling and chaos in favor of trajectories like fluttering and steady descent. This effect has previously been observed in disks with a central hole \cite{vincent2016holes}. Transition from chaos/tumbling to fluttering in these disks was investigated experimentally. As the size of the center hole increases, circulation around the object decreases, resulting in increased trajectory stability. Experiments with falling perforated disks have also demonstrated that lateral displacements of perforated disks are substantially lower than those of other regular systems under the same Reynolds and inertia \cite{zhang2023falling}. This has been confirmed by linear stability analysis and nonlinear simulations \cite{corsi2024instability}.

In this letter, we investigate experimentally the effect of both permeability and porosity on the falling regimes of porous rectangular plates, drawing biological inspiration from porous aerodynamics. We drilled different hole patterns on rectangular plates to modify the porosity and permeability. Our findings confirm the enhanced stability induced by these characteristics, as previously stated: for plates with high inertia values and falling at high Reynolds numbers we do not observe the expected behaviors predicted by the classical phase diagram of impermeable objects. None of our perforated plates performs tumbling, but most of them descend rather fluttering or even steadily. To rationalize the observed fluttering and the fluttering/steady descent transition once the plates are perforated, we postulate that the mechanism causing this behavior is a reduction in circulation around the plate as result of the ability of the throughflow to equalize the pressures across the plate, as already shown in PIV experiments in simpler systems \cite{vincent2016holes}. This hypothesis is tested using phenomenological models \cite{andersen2005unsteady, andersen2005analysis} where the coefficients describing these forces become dependent on porosity and permeability. 

The organization of the manuscript is the following: the description of the experimental setup and methods is detailed in section II. The results of our experimental measures are given in section III and discussed in section IV, before conclusions are drawn in section V.

\section{Material and Methods}

\begin{figure}[ht]
    \centering \includegraphics[scale=0.27]{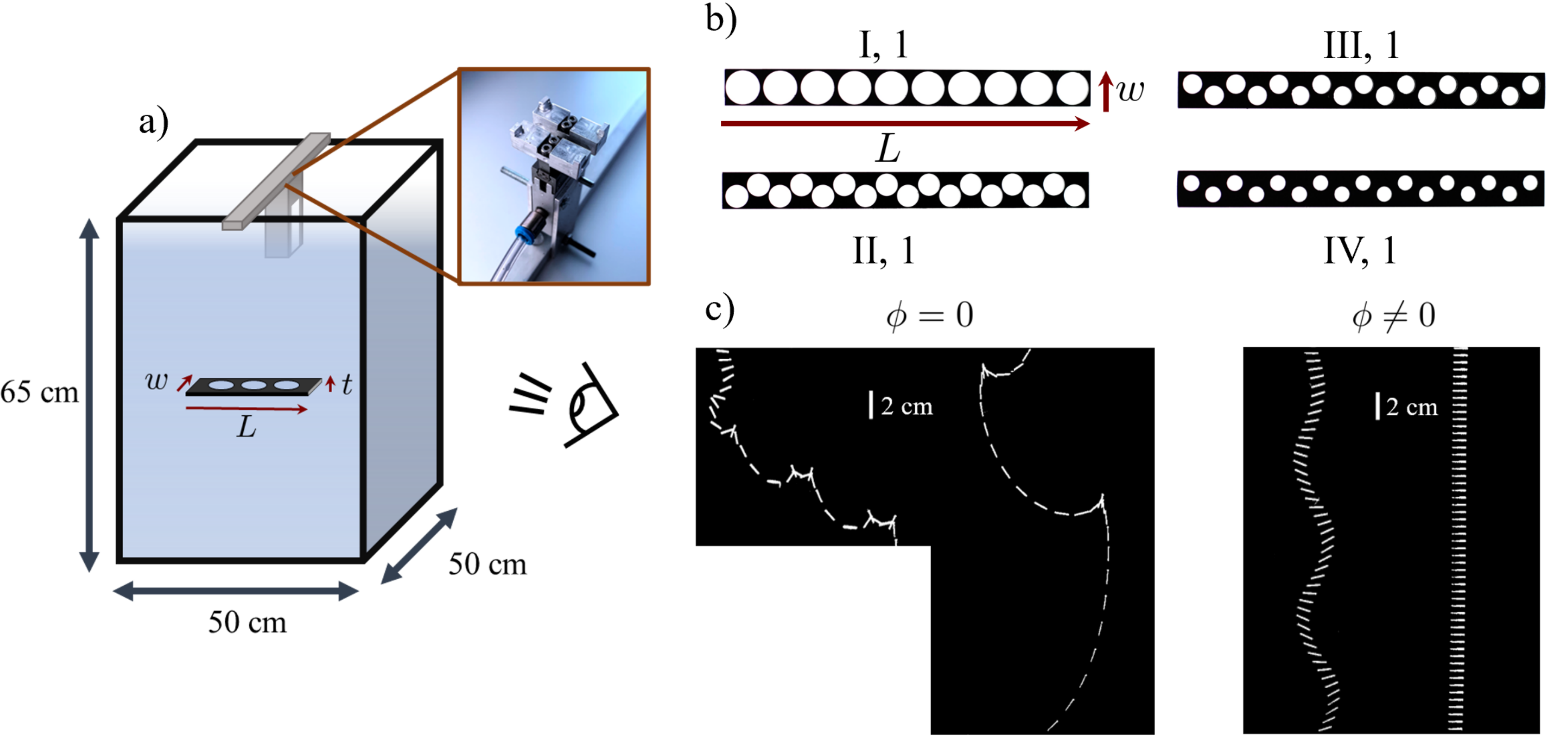}
    \caption{a) Experimental setup with detail of the release mechanism. b) Porous plates released in the experiment. The plates presented correspond to the distributions with the largest holes of each porosity. More details about the porosity patterns and hole distributions can be found in the Supplemental Material \cite{suppmat}.  c) Trajectories of plates released in the experiment. We show two trajectories corresponding to tumbling (porosity V ($\phi_{\mathrm{V}}=0$\%), 1.5 mm aluminum) and fluttering (porosity V ($\phi_{\mathrm{V}}=0$\%), 1 mm aluminum) of impervious plates and two trajectories of fluttering (porosity II ($\phi_{\mathrm{II}}=47$\%), hole distribution 1, 1 mm steel) and steady descent (porosity I ($\phi_{\mathrm{I}}=63$\%), hole distribution 4, 1 mm aluminum) of porous plates. }\label{fig:ExperimentalSetup}
    \end{figure}

The experiments consist in releasing porous plates in a 50x50x65cm aquarium (Olibetta, Aqua Tower 163 Model, Fig. \ref{fig:ExperimentalSetup} a)) filled with water of density $\rho_f = 1000$ kg.m$^{-3}$ and kinematic viscosity $\nu = 10^{-6}$ m$^2$.s$^{-1}$. The release mechanism is comprised of an air compressor, a parallel gripper (FESTO, DHPC-6-A-NO-S-2), an on/off valve (FESTO, HE-1/8-D-MINI) and plastic tubing to connect the three devices. Under a flow of compressed air and the valve in on position, the parallel gripper closes and holds the object. Still with the compressed air flowing, we switch the valve to the off position to let the air out and the gripper immediately opens, releasing the object. The gripper is placed at the top of the aquarium equidistant from the walls, while the lower part of the gripper and the object are already under the water surface  to avoid effects coming from an interaction with the free surface.

The porosity of the plates results from drilling holes into them. All the plates have been produced in the mechanical workshops at EPFL and have been manufactured with the same length, $L=10$ cm, and width, $w=1$ cm. This aspect ratio has been chosen to minimize three-dimensional effects. Since the porosity of the plates considerably reduces their inertia, we decided to produce  plates with different thicknesses and densities to explore a wider range of the parameter space. The choice of plates thicknesses responds basically  to technical limitations to achieve holes without imperfections and to maintain the flat shape of the plate. On the one hand, because the majority of the holes were drilled with drill bits, the plates must be thick enough to resist twisting when the drill bit is used. With the technical tools at our disposal, we can offer the plates a minimum thickness of 1 mm without affecting their shape when the drill penetrates them. On the other hand, excessively thick plates can produce irregular patterns on the plate surfaces, especially for high-density materials. We chose, therefore, to work with three plates thicknesses: $t=$1, 1.5 and 2 mm, and two materials: aluminum ($\rho_{\mathrm{aluminum}} = 2700$ kg.m$^{-3}$) and steel ($\rho_\mathrm{steel} = 7800$ kg.m$^{-3}$). The permeability and porosity are then modified by drilling a variety of porous patterns (see Supplemental Material \cite{suppmat} for further details and Fig. \ref{fig:ExperimentalSetup} b)). Porosity $\phi$ is defined as the ratio between void volume and the total volume of the plate, noting that an impervious (also called impermeable) plate has $\phi=0$. Fig. \ref{fig:ExperimentalSetup} c) shows indeed that the porosity has an effect on the trajectories that falling objects perform. To observe the influence of porosity on the plates falling regimes we have drilled different porosity patterns which are summarized in Table 1. The experimental batch is composed of 78 plates altogether: 18 with $\phi_{\mathrm{I}} \sim 63\%$, 18 with $\phi_{\mathrm{II}} \sim 47\%$, 18 with $\phi_{\mathrm{III}} \sim 33\%$, 18 with $\phi_{\mathrm{IV}} \sim 21\%$ and 6 with $\phi_{\mathrm{V}} = 0$ (impervious plates). In each porosity pattern several hole distributions, varying in size and number to keep $\phi$ constant, are employed to vary the permeability (for additional information about the arrangement, number and size of the holes see the Supp. Mat. \cite{suppmat}). 

\begin{table}[H]
\begin{tabular}{|c|c|c|c|c|c|}
\hline
& \begin{tabular}[c]{@{}c@{}}Porosity pattern I\\ $(\phi_{\mathrm{I}} \sim 63 \%)$\end{tabular} & \begin{tabular}[c]{@{}c@{}}Porosity pattern II\\ $(\phi_{\mathrm{II}} \sim 47 \%)$\end{tabular} & \begin{tabular}[c]{@{}c@{}}Porosity pattern III\\ $(\phi_{\mathrm{III}} \sim 33 \%)$\end{tabular} & \begin{tabular}[c]{@{}c@{}}Porosity pattern IV\\ $(\phi_{\mathrm{IV}} \sim 21 \%)$\end{tabular} & \begin{tabular}[c]{@{}c@{}}Porosity pattern V\\ $(\phi_{\mathrm{V}} = 0 \%)$\end{tabular} \\ \hline
Hole distribution & Rectangular & Hexagonal & Hexagonal& Hexagonal & No holes \\ \hline
Number of plates  & 18 & 18 & 18 & 18 & 6 \\ \hline
$Da$ range    & $4.9\cdot 10^{-4}-1.6\cdot 10^{-5}$ & $2.1\cdot 10^{-4}-2.9\cdot 10^{-5}$  & $1.1\cdot 10^{-4}-1.5\cdot 10^{-5}$  & $5.0\cdot 10^{-5}-6.6\cdot 10^{-6}$  & 0 (Impervious)\\ \hline
\end{tabular}
\caption{Arrangement of the holes, number of plates used in the experiments and range of permeabilities for the different porosity patterns composing the batch.}
\label{tab:porosityPlates}
\end{table}

The permeability $k$ of a plate of thickness $t$ drilled with holes of radius $R$ and porosity $\phi$ reads \cite{pezzulla2020deformation,jensen2014flow}:
\begin{equation*}
    k =\phi  \frac{R t}{3\pi c_{\mathrm{corr}} + \frac{8 t}{R} }.
\end{equation*}
The parameter $c_{\mathrm{corr}} = 1 - p \left( \phi \sqrt{3}/2 \pi  \right)^{3/2} $ expresses the geometrical properties of the porosity pattern through $p = 1.9$ for a square pattern or $p = 2.3$ for a hexagonal lattice \cite{jensen2014flow}. The dimensionless permeability is defined through the Darcy number, expressed as:
\begin{equation*}
    Da = \frac{k}{L w}.
\end{equation*}
Permeability $k$ ranges from $6.6\cdot 10^{-9}$ to $4.9\cdot 10^{-7}$ m$^2$ for our porous plates, allowing us to cover a wide variety of $Da$. For instance, highly porous aluminum foams ($\phi \sim 90 \%$) attain permeabilities around $10^{-7}$ m$^{2}$  \cite{kim2001forced}, which is comparable to the most porous plates employed in this study.

The dimensionless moment of inertia $I^{*}$ and the Reynolds number $Re$ are the other two fundamental dimensionless parameters for the characterization of falling regimes. Following Smith \cite{smith1971autorotating}, $I^{*}$ reads:
\begin{equation*}
    I^{*} = \frac{I}{\pi \rho_f w^4/32},
\end{equation*}
where the denominator of this expression corresponds to the moment of inertia per unit length of a fluid cylinder of density $\rho_f$ and diameter $w$. $I$ is the moment of inertia of the plates calculated using superposition of moments of inertia and parallel axis theorem \cite{marion1988classical}, therefore accounting for the detailed hole pattern. The Reynolds number is defined as:
\begin{equation*}
    Re = \frac{U w}{\nu},
\end{equation*}
with $U$ a typical velocity associated to the fall. The choice of $Re$ as one of the parameters to characterize the falling regimes of the plates is a tricky issue. First, if we define $U$ as the descent velocity $Re$ becomes an \textit{a posteriori} parameter whose value is only known after the experiment is completed. Furthermore, the majority of the plates do not descend steadily but rather follow time-dependent trajectories such as fluttering or tumbling, making it difficult to determine an accurate falling velocity. Other authors have rather used the characteristic velocity obtained by balancing buoyancy-corrected gravity with typical drag force scaling like $\rho_f L w U^2$ \cite{auguste2013falling}, using the Archimedes number $Ar= w \sqrt{2(\rho_{\mathrm{material}}/\rho_f-1) t g}/ \nu$, instead of $Re$. Despite these limitations, the majority of investigations \cite{smith1971autorotating} make use of the time-averaged falling velocity to define the Reynolds number, and we thus decided to use this parameter in our analysis.

The aquarium is illuminated from the top with a 20 watts Easy LED 2.0. Since the observed falling trajectories are essentially two dimensional, the plates are painted in black except for the front edge to enhance contrast and facilitate tracking. Each plate was released four times from rest at zero initial conditions obtaining systematically the same falling trajectory except for three cases that will be discussed in the next section. In these circumstances, we achieved two alternative behaviors equally probably. Therefore, we doubled the number of trials and took extra precautions to eliminate any external disturbances that may affect the plate's release and modify its initial conditions. Following this procedure, the findings were consistent, and the observed behaviors remained uniform. Considering the dimensions of our plates, we confirmed that the height drop is adequate to witness long-term dynamics and there is no secondary transition towards another regime. Velocity signals show two differentiated stages: a transient state up to the first $0.2\ \mathrm{s}$  after dropping, and a steady state with oscillations around a mean value. The falling regimes assessed in this state are identical and the relevant measurable parameters of this steady state hardly differ in the experiments performed. The videos have been recorded with a Sony RX10 IV at a frequency acquisition of 100 fps to prioritize image quality and precision tracking. Image analysis has been performed with ImageJ and data post-processing with Mathematica.

\section{Results}

\begin{figure}[ht]
    \centering \includegraphics[scale=0.85]{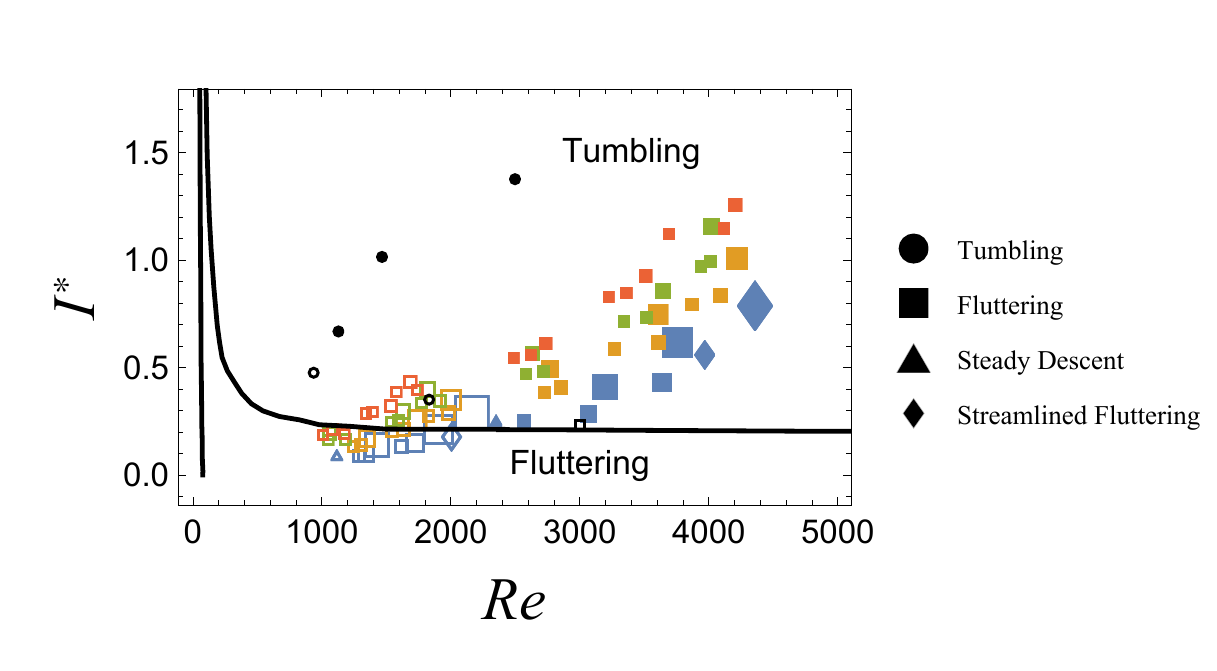}
    \caption{Regime diagram of the plates collected in table \ref{tab:porosityPlates}. Thick black lines represent the separation boundaries between falling regimens according to Smith's phase diagram \cite{smith1971autorotating}. Falling regimes are symbolized by circles (tumbling), squares (fluttering), triangles (steady descent) and diamonds (streamlined fluttering). Plates porosity is represented by color: blue for $\phi_{\mathrm{I}} \sim 63 \%$, orange for $\phi_{\mathrm{II}} \sim 47 \%$, green for $\phi_{\mathrm{III}}\sim 33 \%$, red for $\phi_{\mathrm{IV}} \sim  21 \%$ and black for $\phi_{\mathrm{V}}=0 \%$. Bigger symbols correspond to more permeable plates, symbol size is an affine function of $Da$. Steel and aluminum plates are indicated by full or empty symbols, respectively.}
    \label{fig:PhaseDiagram}
    \end{figure}
    
The plates' falling behaviors are shown and classified in Fig. \ref{fig:PhaseDiagram}. The classical \cite{smith1971autorotating} regime diagram for impervious falling objects $(I^{*}, Re)$ with the borders delimiting the different falling regimes is presented by black lines, together with different symbols indicating our porous plates. The difficulty lies in representing $I^{*}$, $Re$, $\phi$ and $Da$ at the same time in a two-dimensional plot. Since $I^{*}$ and $Re$ are the variables used in the phase diagram, we have labeled porosity by color and permeability by marker size. The colors used are blue for $\phi_{\mathrm{I}} \sim 63 \%$, orange for $\phi_{\mathrm{II}} \sim 47 \%$, green for $\phi_{\mathrm{III}}\sim 33 \%$, red for $\phi_{\mathrm{IV}} \sim  21 \%$ and black for $\phi_{\mathrm{V}}=0 \%$. Marker size is directly correlated with permeability: the larger the markers are, the more permeable the plates. To avoid having a zero-size marker for the impermeable plates, we utilize a minimum marker size for these and for the less permeable plates. The figure clearly distinguishes two sets of porous plates: this division corresponds to the materials aluminum (empty symbols) and steel (full symbols). Last and most important, falling behavior is represented via marker shape. In our experimental study we have found four falling behaviors shown in the figure: tumbling (circles), fluttering (squares), steady descent (triangles) and a fourth behavior we have called streamlined fluttering (diamonds). Robustness of the experimental setup has been verified with the impervious plates, obtaining an excellent agreement with Smith's phase diagram \cite{smith1971autorotating}. Among these six plates, five tumble except for the 1 mm aluminum plate located at the border between the two regimes, which flutters. Unlike other experiments \cite{andersen2005unsteady}, chaos is not observed in our experimental setup. However, it is not our aim to review the behavior of impermeable plates but simply to validate the procedure with the results of the literature. Remaining points in the diagram correspond to the permeable plates with the porosities defined above. As soon as porosity and permeability come into play, the tumbling behavior disappears and the plates become stabilized: the fall occurs in a vertical trajectory with minimal deviations from the release point. Except for five porous plates, all the other 67 flutter.

\begin{figure}[ht]
    \centering \includegraphics[scale=0.85]{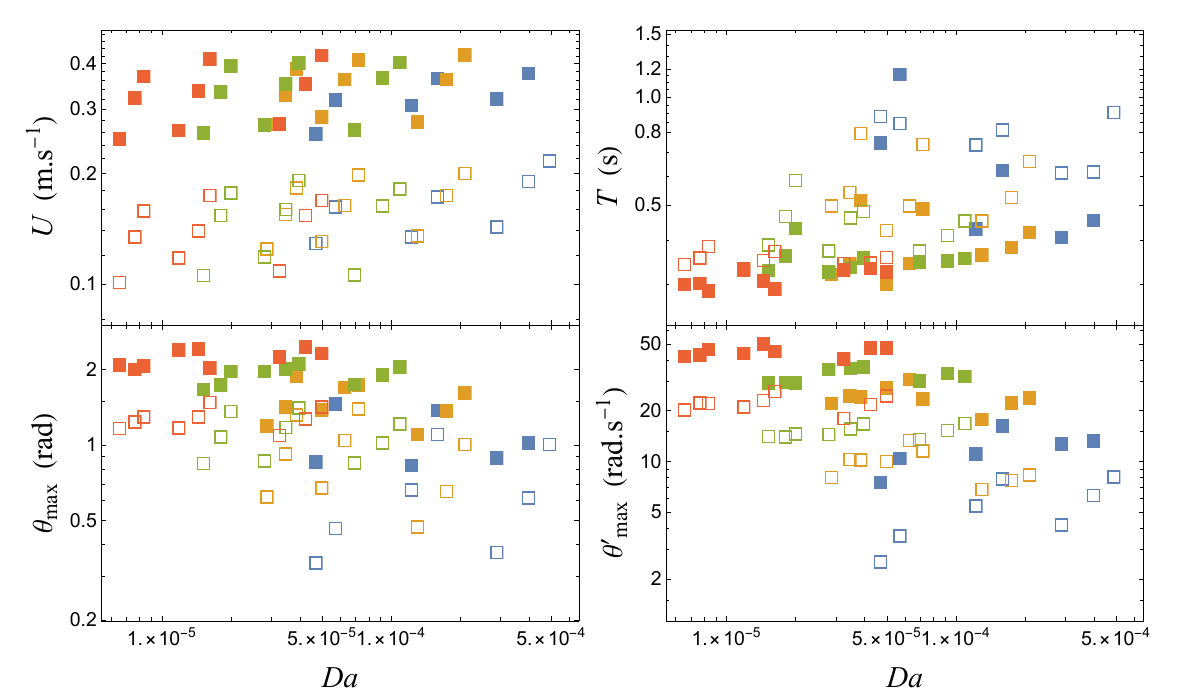}
    \caption{Characteristics of fluttering plates. Falling velocity $U$, period $T$, angular amplitude $\theta_{\mathrm{max}}$ and angular velocity amplitude $\theta'_{\mathrm{max}}$ as a function of Darcy number. Plates porosity is represented by color: blue for $\phi_{\mathrm{I}} \sim 63 \%$, orange for $\phi_{\mathrm{II}} \sim 47 \%$, green for $\phi_{\mathrm{III}}\sim 33 \%$, red for $\phi_{\mathrm{IV}} \sim  21 \%$ and black for $\phi_{\mathrm{V}}=0 \%$. }    \label{fig:FlutterDarcy}
    \end{figure}

To further investigate the effect of permeability and porosity and their influence on fluttering we have identified four experimental variables that describe this falling regime: the average descent velocity $U$, the period of the fluttering oscillations $T$, the amplitude of the oscillations $\theta_{\mathrm{max}}$ and the amplitude of the angular velocity $\theta'_{\mathrm{max}}$. These four magnitudes are shown in Fig. \ref{fig:FlutterDarcy} as a function of the Darcy number. Falling velocity exhibits an appreciable dispersion and no clear trend is observed as a function of the Darcy number but it is mainly led by the dimensionless moment of inertia. That is not the case for the angular characteristics of the motion which seem to be affected by the plates permeability. For a same porosity, $Da$ seems to grow the oscillation period and to decrease both angle and angular velocity amplitudes. Two points of porosity $\phi_{\mathrm{I}}$ (blue) around $Da = 5.5 \cdot 10^{-5}$ appear to be out of the trend, specially for $\theta_{\mathrm{max}}$ and $\theta'_{\mathrm{max}}$. These two points correspond to the aluminum plates of the third hole distribution of porosity $\phi_{\mathrm{I}} $ (I, 3), which have lower values of amplitude and oscillate much slower than the rest of the plates. Low values of these two variables may indicate a proximity to a transition where no fluttering occurs and a steady descent is observed. Flutter characteristics are shown as a function of the plates' inertia in the Supplemental Material \cite{suppmat}.

Although most plates exhibit fluttering in their fall regimes, we have found another behavior that corresponds to the limit of fluttering plates when there is no oscillation: steady descent, represented in Fig. \ref{fig:PhaseDiagram} by triangles. The only plates where this behavior has been observed are two, corresponding to porosity I and hole distribution 4. These two plates are 1 mm thick and have the same holes distribution engraved on them; their only difference is density, since one is made of aluminum and the other of steel, which affects the inertia and the terminal velocities. The experimental ratio between both velocities is approximately $\left( v_{\mathrm{steel}}/v_{\mathrm{aluminum}} \right)_{\mathrm{exp}} \sim 2.2 $, very close to the theoretical value predicted by the balance between drag force and buoyancy-corrected gravity
$ \left(v_{\mathrm{steel}}/v_{\mathrm{aluminum}} \right)_{\mathrm{theo}}  =  \sqrt{ \frac{\rho_{\mathrm{steel}}-\rho_f}{\rho_{\mathrm{aluminum}}-\rho_f  }   }   \sim 2 $. Surprisingly, these plates are those within $\phi_\mathrm{I}$ that have the lowest permeability value and would therefore be expected to more closely resemble the behavior of impermeable plates. We shall interpret this fact in the next section. Actually, these plates with the found values of $I^{*}$ and $Re$ should flutter if they were impervious. Impervious objects performing steady descent is a regime which occurs only for low $Re$ values, less than $Re \sim 80$ for plates \cite{smith1971autorotating} and $Re < 100-150$ for disks \cite{field1997chaotic}. These behaviors have typically been observed in experiments performed in glycerol-water solutions where viscous effects are amplified and can decrease the $Re$ by more than a factor of 10, which contrasts with our experiments where the steady descent occurs for moderate Reynolds values ($>1000$) and certainly much greater than those established in \cite{smith1971autorotating,field1997chaotic}. 

The last observed behavior is represented by diamonds and is the one we have called streamlined fluttering. This falling regime consists of a first stage where the plate modifies its orientation until reaching an angle near $\pi/2$ followed by a second phase where the plate begins to fall with its thickness parallel to the bottom of the aquarium. The plate then oscillates around $\theta = \pi/2$ with a smaller angular amplitude but with a much shorter period and therefore a higher angular velocity (see Fig. 6 in Supplemental Material for details \cite{suppmat}). Not too surprisingly, given the area of the frontal exposed surface, falling velocities before the transition in classical fluttering are smaller than those measured in streamlined fluttering. Furthermore, the characteristic length in the first stage is the width of the plate, and in the second, it is the thickness, so that the Reynolds associated with each of these two phases of the motion should actually not be the same. In Fig. \ref{fig:PhaseDiagram}, however, we have positioned these points according to the Reynolds before the transition. This behavior has been systematically observed in three cases: porosity I, hole distribution 1, 2 mm steel plate; porosity I, hole distribution 2, 2 mm aluminum plate and porosity I, hole distribution 2, 2 mm steel plate. These plates are characterized by a very high porosity ($\phi_{\mathrm{I}}$) and a large thickness. It is curious, however, that this behavior is not observed in porosity I, hole distribution 1, 2 mm aluminum plate but is present in the hole distribution 2. In this case we have repeated each experiment eight times to detect a possible dependence on the initial conditions and the transition to a streamlined fluttering occurs in all experiments. This effect seems to be dominated by moderate inertia $I^{*}$ and large $Re$ and $\phi$ although we do not know if this falling regime is inherent to the experimental setting (influence of lateral walls or approaching bottom during the fall, plate curvature caused by machining) or if there is a more general physical mechanism that triggers it. Streamlined fluttering was also observed in other three cases: porosity I, hole distribution 2, 1.5 mm steel plate; porosity II, hole distribution 2, 2 mm steel plate and porosity III, hole distribution 1, 2 mm steel plate. The problem here was the appearance of two behaviors equally- probably due to the introduction of perturbations at the moment of release. Being thorough and performing the standard procedure, we discovered that the plaques only performed fluttering when (almost) perfectly released.  We ascribed this result to plates with properties that place them in a falling regime approaching a transition in the regime diagram.

\begin{figure}[ht]
    \centering \includegraphics[scale=1.4]{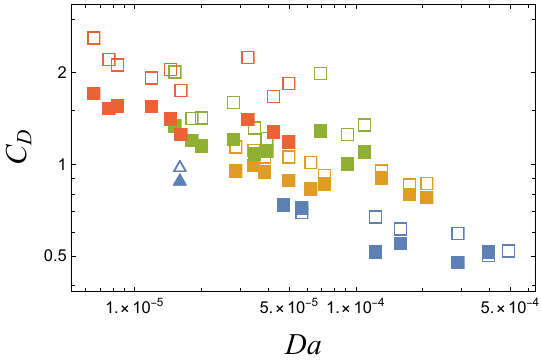}
    \caption{ Drag coefficient as a function of Darcy number.}    \label{fig:DragCoefficient}
    \end{figure}

Additionally, we have calculated the drag coefficients associated to fluttering and steady descent. These coefficients are shown in Fig. \ref{fig:DragCoefficient}. To obtain these values we have balanced buoyancy-corrected gravity with quadratic drag using the mean values of the falling velocity $U$. The trend is consistent with other results in the literature \cite{pezzulla2020deformation} and shows a decrease of the drag coefficient as $Da$ increases. For the same porosity and inertia, coefficients are bigger for aluminum than for steel plates, which can be attributed  to the decreasing trend of drag coefficient with increasing Reynolds number.

\section{Discussion}

The effect of drilling a regular pattern, rectangular or hexagonal, of holes in a plate was shown experimentally to have a stabilizing effect on the falling regime, at least in the limited parameter range explored in this study. This effect, characterized by the total elimination of tumbling, does indeed depend on the number and size of holes; and therefore on the permeability and porosity. It is surprising that plates, even the less porous ones, located in the tumbling region of impervious ones in Fig. \ref{fig:PhaseDiagram} never exhibit a tumbling behavior, unlike the disks with a central hole in the work of \cite{vincent2016holes}, where this falling regime appears for porosities around 20\% and higher. Likewise, extreme behaviors might be expected for the highest permeabilities as these are the opposite extremes of the solid plates. This is, again, not the case for the first porosity pattern ($\phi_I \sim 63 \%$): while the most permeable plates exhibit a well defined fluttering behavior, the less permeable plates adopt a steady descent falling regime. To rationalize these behaviors, we investigated the phenomenological model introduced by \cite{andersen2005analysis} and how the parameters involved may vary with the addition of porosity/permeability. The differential equations that constitute the model read:
\begin{equation*}
    \begin{split}
        (m+m_{11}) v'_{\bar{x}} &= (m+m_{22}) \theta' v_{\bar{y}} - \rho_f \Gamma v_{\bar{y}} - m' g \sin\theta - F_{\bar{x}}^\nu, \\
        (m+m_{22}) v'_{\bar{y}} &= -(m+m_{11}) \theta' v_{\bar{x}} + \rho_f \Gamma v_{\bar{x}} - m' g \cos\theta - F_{\bar{y}}^\nu, \\
        (I+I_a) \theta'' &= (m_{11}-m_{22}) v_{\bar{x}} v_{\bar{y}} - \tau^\nu. \\
    \end{split}
\end{equation*}
See the Supplemental Material \cite{suppmat} for a complete description of the model parameters. The aero-hydrodynamic forces that a falling plate experiences are modeled in a quasi-steady approximation \cite{andersen2005unsteady, andersen2005analysis} and are, in addition to the added mass, lift and drag. The lift force depends mainly on the circulation, which is written in this model as:
\begin{equation*}
     \Gamma = -2 C_T  w \frac{v_{\bar{x}} v_{\bar{y}}}{\sqrt{v_{\bar{x}}^2 + v_{\bar{y}}^2}} + 2 C_R w^2 \theta',
\end{equation*}
where $C_T$ and $C_R$ are the coefficients of the translational and rotational lift, respectively. 

The drag force and dissipative torque read:
\begin{equation*}
\textbf{F}^{\nu} = \rho_f w \left[ A - B \frac{v_{\bar{x}}^2 -v_{\bar{y}}^2}{v_{\bar{x}}^2 + v_{\bar{y}}^2} \right] \sqrt{v_{\bar{x}}^2 + v_{\bar{y}}^2} (v_{\bar{x}},v_{\bar{y}}), \qquad \tau^{\nu} = \pi \rho_f w^4 \left[  \frac{V}{w} \mu_1 + \mu_2 |\theta'| \right] \theta',
\end{equation*}
where $A$ and $B$ are dimensionless constants accounting for the isotropic and anisotropic drag, respectively.  Lastly, $\mu_1$ and $\mu_2$ are the linear and quadratic dimensionless coefficients characterizing the dissipative torque. 

This model therefore employs 6 parameters to explore the fluttering/tumbling transition: the translational and rotational lift coefficients $C_T$ and $C_R$ describe the effect of circulation, $A$ and $B$ quantify the translational drag, and $\mu_1$ and $\mu_2$ describe the dissipative torque. We hypothesize that these parameters, assumed to be constants in the original model, are now functions of $\phi$ and $Da$ for permeable plates. A joint combination of these constants may play a role in the stabilizing effect induced by permeability. Moreover, previous studies \cite{vincent2016holes} have shown an appreciable reduction of circulation in the case of discs with a central hole as opposed to impervious discs. These findings suggest that the falling mode may strongly depend on the value of the circulation around the object. We then performed a thorough analysis on how these two constants $C_T$ and $C_R$ varied as porosity and permeability changed. This analysis relies on two further simplifications: $B$ and $\mu_1$ are set to 0 since they do not prevent tumbling-fluttering nor fluttering-steady descent transition and are not the predominant source of drag and friction torque in the range of Reynolds number measured in the experiment. The coefficient $\mu_2$ for the dissipative torque may be related to $A$ using scaling laws on the drag so that $\mu_2 \sim A $ (Supp. Mat. \cite{suppmat}). Besides, $A$ and $\mu_2$ are obtained from actual measures of the drag coefficients (Fig. \ref{fig:DragCoefficient}). Balancing the drag force with the buoyancy-corrected gravity when the plate is falling leads to $2A =C_D$. 

 \begin{figure}[ht]
    \centering \includegraphics[scale=0.28]{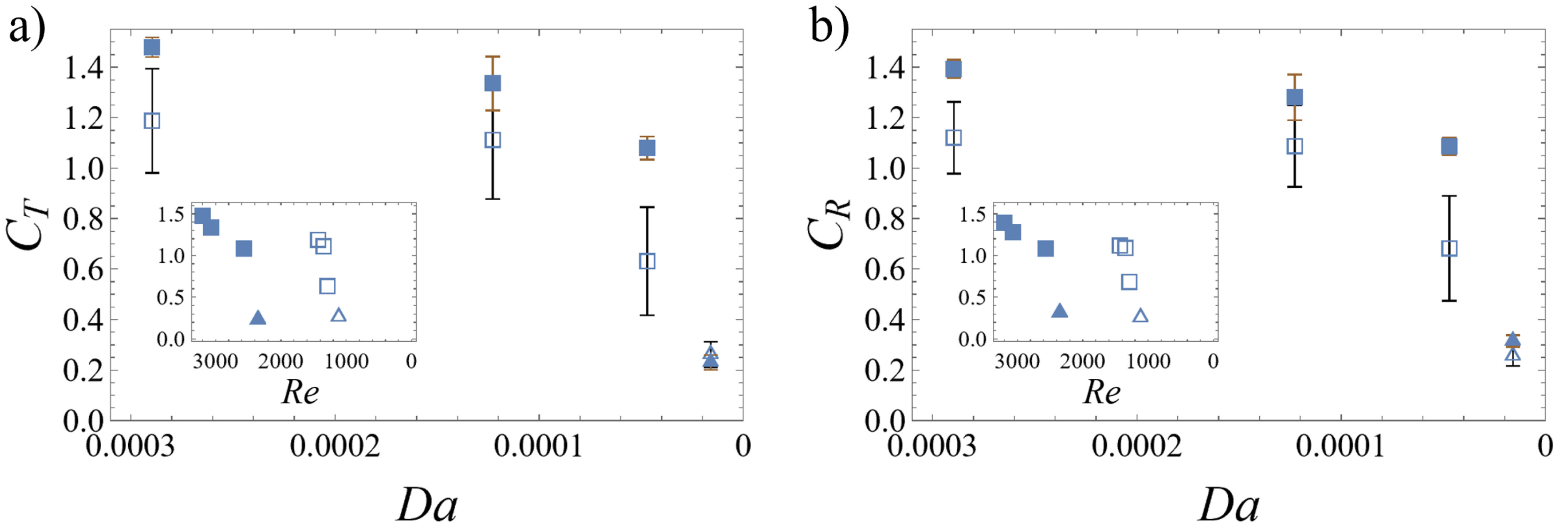}
    \caption{a) $C_T$ and b) $C_R$ coefficients as a function of Darcy Number for the 1 mm aluminum (empty symbols) and steel (full symbols) rectangular pattern plates. Black and brown bars correspond to the errors associated to aluminum and steel plates, respectively. In insets, $C_T$ and $C_R$ as a function of Reynolds number.    }    \label{fig:CtCrPlates}
    \end{figure}

To confirm our hypothesis that the steady fall/fluttering transition is accompanied by a diminution of the circulation and, hence, of the coefficients $C_T$ and $C_R$, we determined these coefficients for the 1 mm aluminum and steel plates belonging to the porosity pattern I. These plates start a fluttering behavior that gradually attenuates as the period of their oscillations increases and their angular velocity decreases until they fall steadily as permeability decreases. We show in Fig. \ref{fig:CtCrPlates} the behavior of $C_T$ and $C_R$ as a function of $Da$ for the 1 mm aluminum and steel plates. The values of $C_T$ and $C_R$ have been obtained using a SVR algorithm \cite{scikitlearn} (additional details can be found in the Supplemental Material \cite{suppmat}) aiming at best capturing the steady fall/fluttering transition.  We do not intend to propose a numerical technique for calculating these coefficients, but rather to demonstrate the qualitative dependence they exhibit. Nevertheless, the values we found for the less permeable plates are consistent and in the same order of magnitude as those observed in previous studies \cite{andersen2005unsteady}. In this figure we corroborate our main hypothesis about the stability mechanism. The coefficients characterizing the circulation around the object decrease as the pattern of holes varies, i.e. reducing the permeability while keeping the porosity constant. Both values are relatively stable for the first two hole distributions but decrease sharply to values about 5 times lower for the less permeable plates. In this case the decrease in circulation is truly appreciable and is likely at the origin of fluttering-steady descent transition that we observed in the experiments. However, in other plate configurations, such a significant decrease is not observed, preventing us from concluding on a general trend for all plates at constant porosity. While inertia does not play any role in the values of the coefficients, $Re$ may have a significant impact, at least in the fluttering regime. $Re$ is approximately twice as large for steel plates as for aluminum plates in Fig. \ref{fig:CtCrPlates}, regardless of the $Da$ value. In fluttering, both $C_T$ and $C_R$ values are consistently greater for steel than for aluminum; however, these coefficients appear to be unaffected by Reynolds number when the plates fall steadily.

\section{Conclusions}

We have presented in this paper an experimental result of stability induced by porosity and permeability. We discovered that by drilling different porosity patterns, which differ in their arrangement, number of holes, and hole size, we can avoid tumbling and chaotic behavior in plates that, due to their inertia and Reynolds values, should tumble while falling according to the regime diagram of impervious plates. The majority of the plates flutter, and a few even descend steadily. Furthermore, we discovered another behavior, streamlined fluttering, only in very porous and thick plates, which should be investigated further to determine if the observed behavior is an experimental artifact or if there exists an underlying physical mechanism. We have used a phenomenological model to fit relevant parameters during plate fall and observed how the decrease in circulation emerged as a plausible explanation for the stabilizing mechanism  associated to the introduced porosity. This opens many theoretical and numerical perspectives to explain the circulation generation around permeable plates. Furthermore, additional studies on the spatial distribution of the holes (whether more concentrated at the edges or in the interior) or on non-regular patterns may advance our understanding in the stabilizing effect of porosity and permeability. We believe that these results may be relevant for time-dependent morphing applications \cite{marzin2022shape}, for example in the design of objects that can vary their porosity and permeability while falling, providing better stability.

\section*{Acknowledgements}
The authors want to thank the Workshop at the Institute of Mechanical Engineering (ATME), the Materials Institute workshop (ATMX), and the Workshop of the Institute of Production and Robotics (ATPR) at EPFL for useful suggestions and fabrication of the plates. The authors also acknowledge the EPFL Hub for Image Analysis for helpful advice about the image analysis of the falling plates. Anthony Pieper is thanked for participating in the experiments. JSR acknowledges the support
of the Ministerio de Ciencia, Innovación y Universidades
of Spain under a Margarita Salas contract funded by
the European Union-NextGenerationEU.








\bibliography{porousplates}

\end{document}